\documentclass[12pt,preprint]{aastex}

\shorttitle{Near-UV objects in the CDF-S}
\shortauthors{de Mello et al.}

\begin{document}

\title{The Nature of Near-UV Selected Objects in the Chandra Deep Field South
\footnote{Based on observations taken with the NASA/ESA Hubble Space Telescope,
which is operated by the Association of Universities for Research in Astronomy,
Inc. (AURA) under NASA contract NAS5-26555 and on observations
collected at the European Southern Observatory, Chile}}

\author{D.F. de Mello\altaffilmark{1,2,3}, Jonathan. P. Gardner\altaffilmark{1}, 
T. Dahlen\altaffilmark{4}, C.J. Conselice\altaffilmark{5}, N.A. Grogin\altaffilmark{3} \& A.M. Koekemoer\altaffilmark{4}}

\altaffiltext{1}{Laboratory for Astronomy and Solar Physics, Code 681, Goddard Space Flight Center, Greenbelt, MD
20771}
\altaffiltext{2}{Catholic University of America Washington, DC 20064}

\altaffiltext{3}{Johns Hopkins University, Baltimore, MD 21218}

\altaffiltext{4}{Space Telescope Science Institute, Baltimore, MD 21218}

\altaffiltext{5}{California Institute of Technology, Pasadena, CA 91125}

\begin{abstract}

We present the first results of the pure parallel near-UV observations
with WFPC2 in the Chandra Deep Field South. The main goal is to
search for the population of objects that contribute to the rise
in the volume-averaged star formation rate at intermediate redshifts
(z$<$1). A total of 34 objects, were identified in the WFPC2 images
and their counterparts found in the ACS images. We use template
fitting of the spectral energy distributions to classify the objects
and found that 17 are starbursts, 9 are late-type, 2 are early-type
galaxies, and 6 are stars; two of the late-types and one of the
starbursts are X-ray sources. The colors of the starbursts were
reproduced by the stellar evolutionary synthesis code Starburst99
which shows a mixed population of bursts with ages $<$1 Gyr. Analysis
of the light concentration, asymmetry and clumpiness, shows that
this sample is a mixed bag, containing dwarf ellipticals, early
and late-spirals and peculiar objects which resemble mergers in
progress.

\end{abstract}


\section{Introduction}

A major goal in modern extragalactic astronomy is to understand
how the Universe of galaxies has evolved with cosmic time. The rise
in the volume-averaged star-formation rate at intermediate redshifts
(0$<$z$<$1) is often attributed to a disk epoch when gaseous galactic
disks are converting into stars (e.g. \citet{mad96, roc98, lil98}).
The fate of these galaxies is unknown. They could either fade away
and would not be seen at z=0 or merge to form larger systems. In
either case, they are experiencing strong star formation at
intermediate-z and going through an important phase in galaxy
evolution when star formation rate was higher than that for the
local universe. Little is known of the nature of these objects,
since deep, high resolution images are required to analyze their
properties. Most of what is known comes from the Hubble Deep Field
project \citep{wil96, wil00} which has detected thousands of galaxies
at intermediate to high redshifts. However, the small sizes of the
fields translates to a relatively small effective volume at z$<$1,
limiting the ability to probe the evolution of UV-bright objects
at moderate redshifts. The Advanced Camera for Surveys (ACS) aboard
HST has started to improve this situation covering a larger
field-of-view than WFPC2 with higher resolution.

Ultraviolet (UV) light from hot, high-mass stars dominates the
spectral energy distribution of galaxies going through strong star
formation, also known as starbursts. The UV light can be observed
directly from space or indirectly in the infrared as the dust
reprocesses the UV starlight and re-emits it in the infrared. The
UV observations probe unobscured star formation (see \citet{win02}
for an HST survey of mid-UV morphology of nearby galaxies), therefore,
observing in the UV directly accesses the population of objects
that contribute to the rise in the star-formation rate at
intermediate-z. Knowing the properties of these UV sources is also
important when analyzing high-z galaxies because optical observations
of high-redshift galaxies measure the rest-frame UV. In this Letter
we report the first results of a survey designed to study the the
UV sources by using the HST/WFPC2 (F300W) in combination with the
HST/ACS multi-wavelength images.

\section{Observations and Reduction}

We implemented a WFPC2 pure parallel program aimed at maximizing
the synergy between the parallels and the prime proposals of HST.
For parallel fields that fell within the Great Observatories Origins
Deep Survey (GOODS) fields, we took near-UV images with the F300W
filter (U band) of galaxies that have been imaged at redder
wavelengths with ACS. More details regarding the ACS data are given
in Giavalisco et al. (2003) in this issue.

\begin{figure}
\plotone{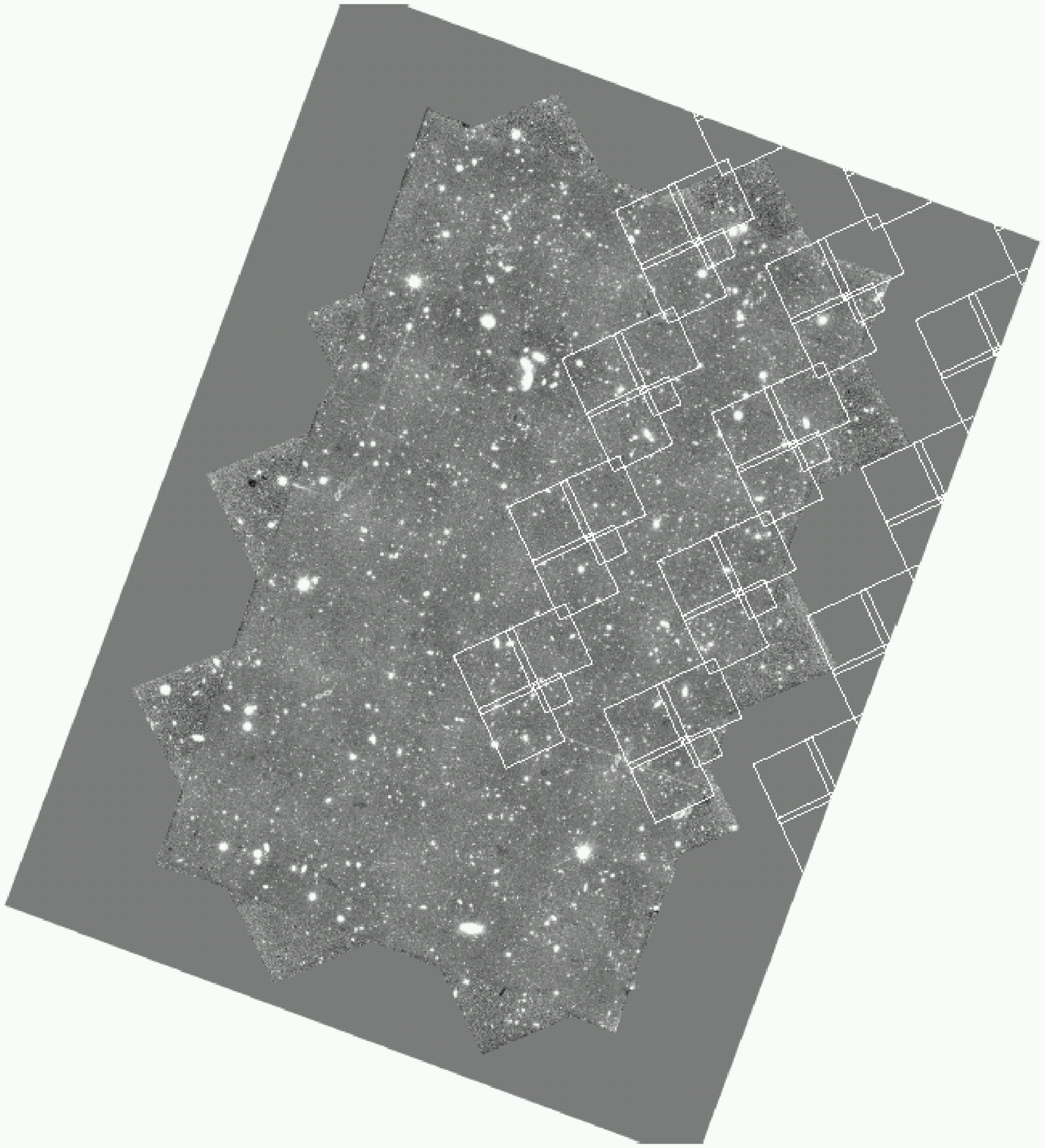}
\caption{WFPC2 fields (white contours) over ACS image of the CDF-S \label{f1}.}
\end{figure}

In this Letter we present preliminary results from this program,
and discuss the sample that was selected from the WFPC2 data taken
during HST/ACS first visit of the Chandra Deep Field South.
Fig.\ref{f1} shows the area observed on top of the ACS mosaic and
the 8 WFPC2 images which were analyzed. The program continued
through the later epochs and in the northern field, and those data
will be discussed in future papers.

WFPC2 images were reduced using the HST pipeline and dithered using
the package PyDrizzle which provides an automated method for
dither-combining and distortion-correcting images. The quality of
the drizzled image was checked by analyzing the PSF of a few stars
present in some of the images.

\section{Sample Selection and Gallery of Objects}

We detected sources on the U-band image using SExtractor v2.2.2
(\citet{ber96}, hereafter SE). Our detection criterion was that a
source must exceed 1.5 $\sigma$$_{sky}$ threshold in 5 contiguous
pixels and we also use a detection filter with a Gaussian FWHM of
4 pixels. A total of 58 objects were identified by SE. Their
magnitude range is 18$<$m$_{U}$$<$26 where m$_{U}$ (AB system) is
SE's mag$_{-}$auto, which is calculated using a flexible elliptical
aperture around every detected object. These numbers agree with
what is expected with the U band (6 objects in  2 to 3-orbit
exposure). For comparison, the HDF project included a long (45
orbit) U exposure and has 133 objects with 23.5$<$m$_{U}$$<$25
(Vega system, \citet{met01}).

The next step was to match the U-band catalog with the ACS B-band
catalog (B$_{435}$) produced by the GOODS team. We adopted a maximum
offset radius of 1.5 arcsec between the WFPC2 coordinates and the
ACS B band, and identified a total of 34 objects. The remaining
objects are either spurious detections or outside the ACS field.
Most of the 34 objects identified are in 4 of the WFPC2 images with
larger exposure times (4320s, 6320s, 7420s, 7620s). From visual
inspection we were able to identify 7 point-like objects in the
sample which were further investigated using the SIMBAD database.
SIMBAD finds at least one star within 10 arcsec of six of the
point-like objects and one QSO. The six point-like objects identified
as stars were removed from the analysis. In Section 5 we discuss
the object identified as a QSO in more detail.

We constructed a gallery of images including the WFPC2 U band, and
ACS B, V, i, z images which is available on-line 
\footnote{http://goods.gsfc.nasa.gov/goods/duilia/gallery/}. The gallery also
includes other parameters such as coordinates, links to the SIMBAD
identifications, spectral types, structural parameters and photometric
redshifts which are described below.

\section{Photometric Redshifts, Spectral Types}

\begin{figure}
\plotone{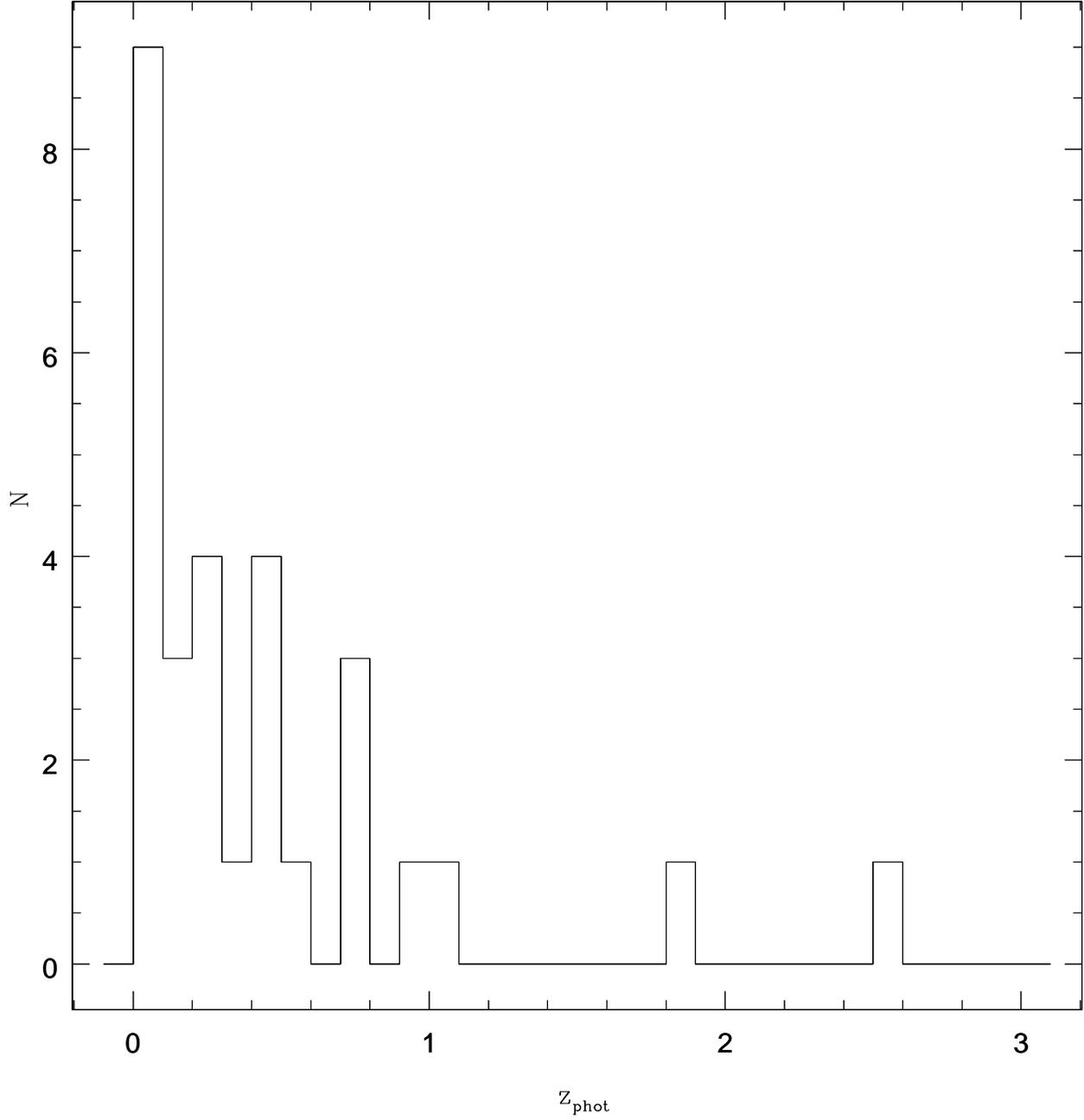}
\caption{Distribution of photometric redshifts. Objects with z$>$ 1 have large uncertainties, see Section 4 for details.
\label{f2}}
\end{figure}

Photometric redshifts were estimated by the GOODS team using the
Bayesian photometric redshift method \citep{ben00} which includes
templates of the spectral energy distribution of E, Sbc, Scd, Im
\citep{col80}, and two starbursts \citep{kin96} -- below we refer
to these as spectral types 1 to 6. The uncertainties estimated from
comparison with ground based spectroscopy are
(${z_{phot}-z_{spec})/(1+z_{spec}})\sim 0.1$. The redshift distribution
of the 28 objects in the U catalog peaks at redshift $<$ 1 (Fig.
\ref{f2}). The WFPC2 data were not included in the photometric
redshift estimation due to large photometric errors in comparison
to ACS photometry.

Aperture magnitudes (diameter=2$''$) in the HST filter system were
calculated using SE. These were transformed into absolute magnitudes
in the Johnson (1966) filter system by adding distance modulus and
appropriate K-corrections. The K-corrections were calculated using
information on filter transmission curves for both systems, spectral
shape of the object (i.e., the best-fitting spectral energy
distribution) and the photometric redshifts. Absolute magnitudes
are obtained using the cosmological model h=0.7, $\Omega_{\Lambda}$=0.7,
$\Omega_{M}$=0.3.

\section{Starbursts and AGNs}

We investigated the morphology and colors of the sample. The spectral
types obtained from the photometric-z estimation were used to
separate the sample into starbursts and non-starbursts. Starbursts
have SED similar to the starbursts in Kinney et al. (1996). In this
Letter galaxies called starbursts are unobscured by dust, however
the non-starburst sample may include dust obscured starbursts.

\begin{figure}
\plotone{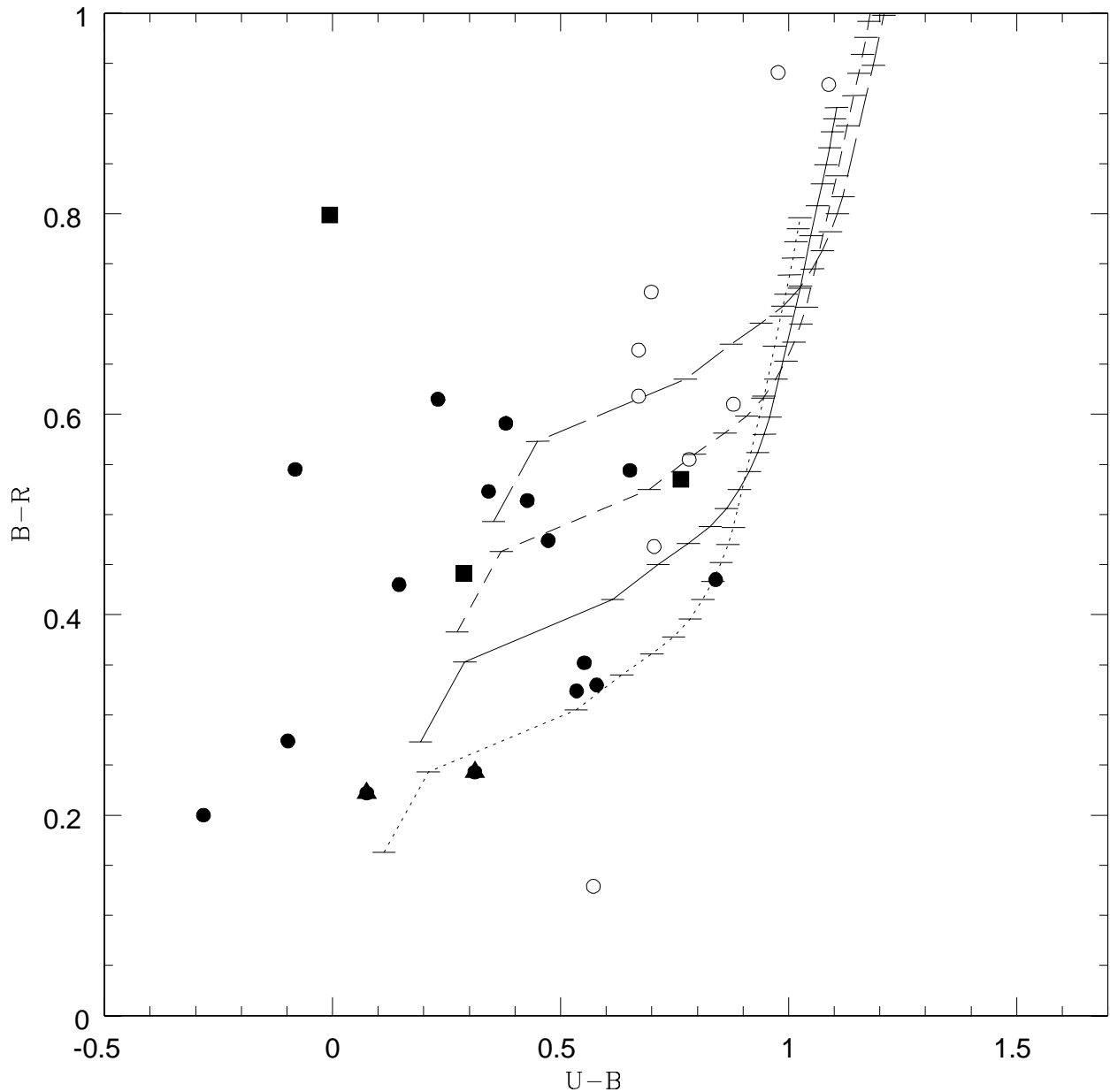}
\caption{Color-color diagram U-B versus B-R, solid symbols are for
the spectral types $>$ 4, X-ray sources marked with squares, high-z
galaxies marked with triangles. Color evolution of starburst galaxies
generated with Starburst99 with continuous star formation,
SFR=1M${_\odot}$/yr and solar metallicity with E(B-V)=0.2, 0.25,
0.3, 0.35 (from bottom to top). Time steps are marked as follows:
t=1 Myr, 10 Myr, 100-900 Myr ($\Delta$t=100 Myr), 1-10 Gyr ($\Delta$t=1
Gyr). \label{f3}}
\end{figure}

In Fig.\ref{f3} we have separated the starbursts from non-starbursts
(filled symbols are galaxies with spectral types $>$ 4, i.e.
starbursts) and have added starbursts colors generated with the
evolutionary synthesis code Starburst99 (\citet{lei99}, see also
\citet{dem00}). Models of the emission from starburst stellar
populations were calculated as a function of age, with a Salpeter
initial mass function (IMF), and stellar evolutionary tracks of
solar metallicity. We used a continuous star formation law at a
constant rate (1 solar mass per year) over 10 Gyr. Extinction of
E(B-V)=0.2, 0.25, 0.3, 0.35 and a Calzetti law \citep{cal00} was
applied to the model. Dust absorption reddens B-R more than U-B
and a moderate extinction correction is able to reproduce the colors
of most galaxies. Starbursts are closer to the young ages ($<1$
Gyr) of the models whereas non-starbursts have colors compatible
with older population; except for one object which has a very low
B-R. This object, although classified as Scd, is a peculiar galaxy
with a double nucleus. Peculiar galaxies have a greater spread of
values in two-color plots than morphologically normal galaxies due
to the great diversity in their star formation history \citet{lar78}.
We have not made any attempt to correct the colors of the models
for nebular line emission. Models including stellar light, nebular
continuum emission, and nebular line emission have been shown in
\citet{joh99}, where it is clear that some of the broadband colors
are strongly affected at ages $<$ 100 Myr.

We investigated the 3 objects which are much bluer (U-B$<$0) than
the models in more detail. These objects have very low photometric
redshifts (z$\sim$0.04) with high uncertainties and could be in
the range 0$<$z$<$0.2. A higher redshift could shift their colors
by at least $\sim$0.3 which will make their colors more typical of
starburst galaxies (see morphology section for more details on
these objects). The bluest non-starburst galaxy is a very peculiar
late-type galaxy and it is probably an obscured starburst.

We also checked the two galaxies with photometric redshifts of
z=1.9 and 2.6. The uncertainties in the photometric redshift of
the first object is extremely high and there is almost no constraint
in the phot-z determination. The 95\% confidence interval is between
z=0.01 and 2.28. The second object has also high uncertainty in
the phot-z but not as high as the previous one with 95\% confidence
interval of having 1.95$<$z$<$3.07. Although their colors are not
used in the analysis, they were kept in the sample and flagged as
triangles in all plots.

We also searched for AGN in the sample. Three objects in the U-band
sample have X-ray emission (see Koekemoer et al. 2003 in this
issue). We adopted 3.0 arcsec as the maximum offset radius between
the U and the ACS z-band coordinates. These objects are flagged as
squares in all plots and are shown in the gallery. Visual inspection
of the three X-ray sources shows that one of them is a point-like
object and the other two are late-type galaxies. The point-like
object is one of the most powerful CXO sources \citep{gia02} in
the entire ACS field (Soft X-ray flux=4.00e-14 $\pm$ 4.69e-16 and
Hard X-ray flux 8.06e-14 $\pm$ 1.65e-15), it is a TypeI AGN
\citep{sch01}. The other two objects are extended and have total
X-ray (0.2-8keV) fluxes 1.77E-16 and 2.66E-16 erg/s/cm$^{2}$ which
are three orders of magnitude lower than the TypeI AGN (1.15E-13).
The object with the lowest X-ray flux has a spectral type typical
of starbursts and peculiar late-type morphology. The third object
has a spectral type similar to intermediate-late spirals.

\section{Morphology}

\begin{figure}
\plotone{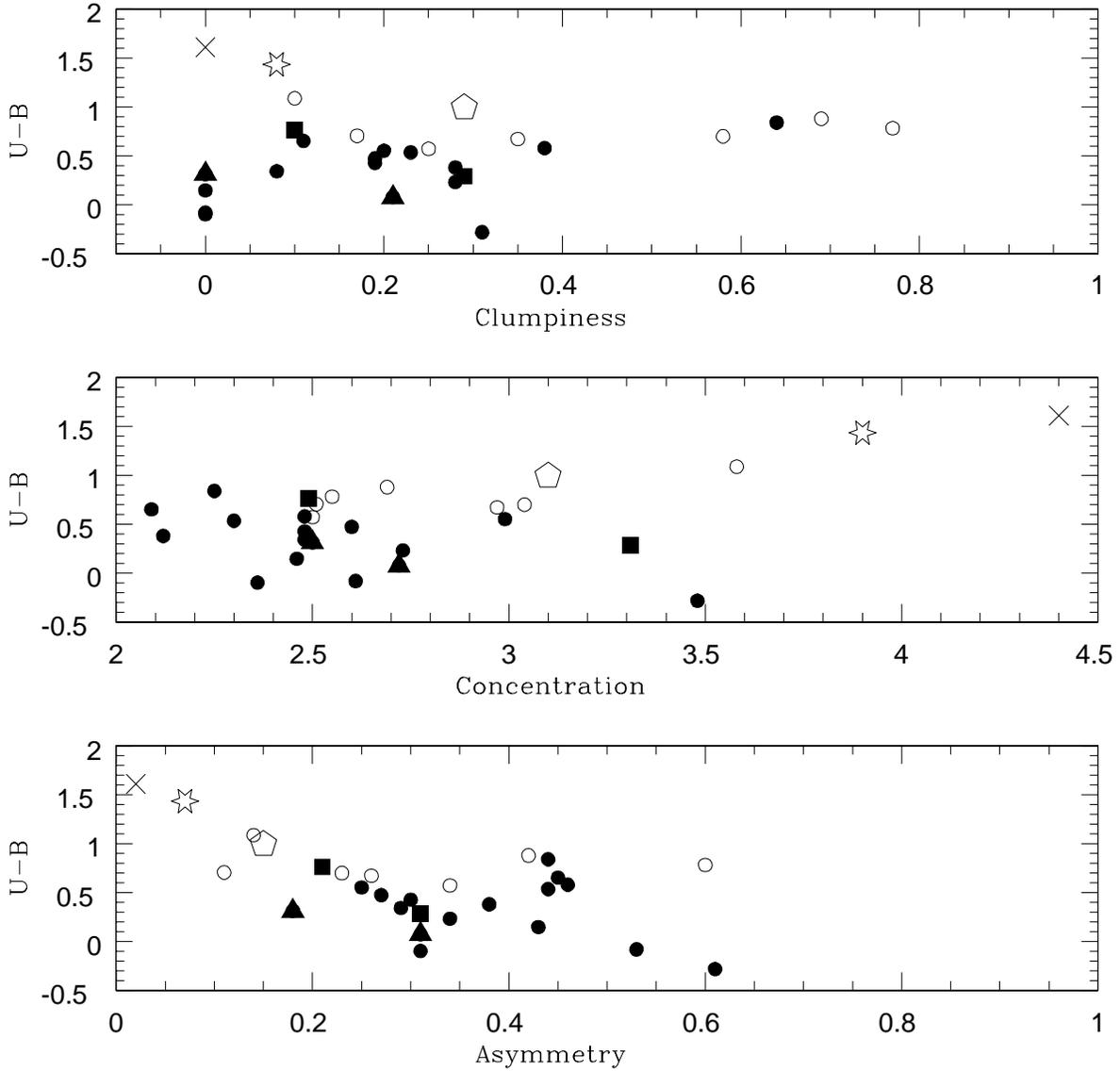}
\caption{Color-asymmetry (bottom), color-concentration (middle)
and color-clumpiness (top) diagrams. X-ray sources are marked with
squares (the star-like X-ray source is not included), high-z galaxies
are marked with triangles. Filled circles are starbursts, diagonal
cross represents typical value of an elliptical galaxy, star
represents Sab and pentagon represents Scd. \label{f4}}
\end{figure}

Galaxy morphology was evaluated by measuring the light concentration
(C), asymmetry (A) and clumpiness (S) of the ACS images in the
rest-frame B band using the method by \citet{con03}, hereafter CAS. 
Early-type galaxies are expected to have a high
concentration index and low asymmetries whereas late-type (disks)
galaxies have low light concentration and high asymmetries. As
suggested by \citet{con00}, a correlation between asymmetry and
color can be seen for most of Hubble types and outliers which are
too asymmetric for their colors are probably dynamically disturbed.
This is true for the near-UV sample, as evident in Fig.\ref{f4}.
The non-starburst galaxy (spectral type 1.67) with the highest
concentration index is also the reddest and the least asymmetric.
The non-starburst galaxy (spectral type 3.33) with the highest
asymmetry value is a spiral galaxy showing tidal effects. The two
starburst galaxies with the highest asymmetry values are also
peculiar, but are in a more advanced stage of interaction than the
previous one. In Fig.\ref{f4} we plot the typical values of U--B
for templates of E (diagonal cross), Sab (star), and Scd (pentagon)
\citep{liu98} and CAS parameters for the same types \citep{con03}.
The CAS values of the templates should be taken with caution since
Conselice uses R-band to estimate them and we are using rest-frame
B band. However, the differences between the CAS values using R
and B band are small ($\Delta$C=0.12 and $\Delta$A, S=0.03). As it
is clear from these figures, the near-UV sample falls in a different
area of the plot with the great majority being bluer, having lower
concentration indexes and higher asymmetries than the templates.
The asymmetry values are typical of starbursts galaxies and ULIRGS
(0.53$\pm$0.22 and 0.32$\pm$0.19, \citet{con03}). A third parameter
obtained with CAS measures the patchiness of the light distribution
(a.k.a. clumpiness) and is shown in Fig. \ref{f4} top diagram. Most
of the galaxies have clumpiness similar to the templates. The three
objects with S$>$0.6 have distorted morphologies with A$>$0.4.

The three parameters together are good evaluators of the morphologies
and were used to investigate the three blue outliers mentioned in
the previous Section. Two of them have C and S values typical of
dwarf ellipticals (C=2.4, 2.6 and S=0.0) and are as faint as dwarf
galaxies (M$_{\rm B}$=--12.4 and --13.9, IDs 1537 and 764 in the
gallery). Visual inspection of these objects confirm these results.
The third blue object has higher concentration, asymmetry and
clumpiness (3.48, 0.61, 0.31, respectively, ID 2232 in the gallery)
than the previous ones. Visually it looks like a merger in progress
of two small nuclei in close contact and a third small knot separated
by 6 pixels from the brightest nucleus. These examples show the
diversity of objects that we are covering in this sample.

We investigated how the CAS parameters of the near-UV sample compares
with the values for all galaxies identified in the ACS field (see
Mobasher et al. in this issue). There are 1134 galaxies with R$_{\rm
AB}$ $<$ 24 within 0.24$<$ z $<$ 1.3 of which $\sim$65\% have SEDs
typical of late-types (C=2.65 and A=0.27), 21\% of early-types
(C=3.25, A=0.23) and 14 \% of starbursts (C=2.53, A=0.33). The
average concentration index of CAS values of the near-UV sample
are C=2.63$\pm$0.35 and A=0.35 $\pm$0.13. The near-UV selected
sample is more asymmetric than the average and even the non-starburst
of the near-UV sample have higher asymmetry (A=0.30 $\pm$0.17),
although their concentration indexes are more typical of earlier types.

\section{Conclusions}

We presented a near-UV selected sample of 34 objects which were
identified in 8 WFPC2 (F300W) images covering part of the GOODS
CDF-S field. Counterparts of all objects were found in the ACS
multi-waveband mosaics which were used to classify them according
to their spectral energy distributions. A total of 17 were classified
as starbursts (1 X-ray source), 9 as late-types (2 X-ray sources) and
2 as early-types. Six stars were removed from the sample. The colors
of the starbursts which were reproduced by the stellar evolutionary
synthesis code Starburst99 are typical of a mixed population of
bursts with ages $<$1 Gyr. Light concentration, asymmetry and
clumpiness were used to analyze the morphology of the UV selected
sample. The majority of the galaxies have lower values of light
concentration and higher asymmetries than typical early and late-type
galaxies. Based on the morphologies we conclude that near-UV sources
at intermediate-z originate in a variety of objects: dwarf ellipticals,
early- and late-spirals and peculiar objects which resemble mergers
in progress. Intermediate redshifts are an important epoch when
the rise in the volume-averaged star formation rate occurs. The
population of objects that contributes to this rise is not homogeneous,
but made of galaxies of different types experiencing strong
star-formation.

\acknowledgments

We are grateful to Benne Holwerda for helping with SExtractor
issues, Warren Hack for helping with PyDrizzle and to the GOODS team. 
Support for this work was provided by NASA through grants GO09583.01-96A
and GO09481.01-A from the Space Telescope Science Institute, which
is operated by the Association of Universities for Research in
Astronomy, under NASA contract NAS5-26555. This research has made
use of the SIMBAD database, operated at CDS, Strasbourg, France.

\end{document}